\documentstyle[nato]{crckapb}
\def\ba{\begin{eqnarray}}\def\ea{\end{eqnarray}}
\def\nn{\nonumber\\}
\def\bc{\begin{center}}\def\ec{\end{center}}
\begin{opening}
\title{Ultraperipheral photoproduction of vector mesons in the nuclear
Coulomb field and the size of neutral vector mesons}
\author{S.R.~Gevorkyan$^{1,2}$, I.P.~Ivanov $^{3,4}$ and N.N.~Nikolaev
$^{3,5}$}
\institute{$^{1}$Joint Institute of Nuclear Research,141980 Dubna,Russia\\
$^{2}$ Yerevan Physics Institute, 375036, Yerevan, Armenia\\
$^{3}$ IKP, Forschungszentrum J\"ulich,D-52425 J\"ulich, Germany\\
$^{4}$ Sobolev Institute of Mathematics, 630090, Novosibirsk, Russia\\
$^{5}$ L.~D.~Landau Institute for Theoretical Physics, 142432
Chernogolovka, Russia}
\end{opening}
\begin{document}
\runningtitle{THE CRCKAPB STYLE FILE}

\begin{abstract}
We point out a significance of ultraperipheral photoproduction of
vector mesons in the Coulomb field of nuclei as a means of measuring
the radius of the neutral vector meson. This new contribution
to the production amplitude is very small compared to the conventional
diffractive amplitude, but because of large impact parameters
inherent to the ultraperipheral Coulomb mechanism its impact on
the diffraction slope is substantial. We predict appreciable
and strongly energy dependent increase of the diffraction slope towards
very small momentum transfer.The magnitude of the effect is proportional to
the mean radius squared of the vector meson and is within the reach of
high precision photoproduction experiments, which gives a unique
experimental handle on the size of vector mesons.
%\vspace*{2cm}\\
\end{abstract}

\section{Introduction}

It was noticed long time ago \cite{P}, that photoproduction of
pseudoscalar mesons in the Coulomb field of nuclei allows one to measure the
radiative width of mesons ($\pi^0 \to \gamma\gamma, ~ V\to P\gamma $
etc.,for the review and references see \cite{L}). The principal
point is that the Primakoff amplitude can be isolated for the presence
of the Coulomb pole in the production amplitude, so that the contribution to
the differential cross section of the process $\gamma+A \to P+A, P=(\pi^0,
\eta,\eta\prime) $ from one-photon Coulomb exchange reads
\ba
{d\sigma\over dq^2}={8\pi\alpha_{em} Z^2\over {m_P}^3}
\Gamma_P{q^2\over(q^2+\Delta^2)^2} F_A^2(\vec{q})\, .
\label{eq:1}
\ea
Here $m_P$ is the mass of the meson, $Z$  is the nucleus charge
number; $\alpha_{em}= {e^2\over 4\pi}$ is the fine structure
constant, $ q=\mid\vec q\mid $ and
\ba
\Delta={m_P^2\over 2E_{\gamma}} \label{eq:2}
\ea
are the transverse and the longitudinal momentum transfer,
respectively, $ F_A(\vec{q})$ is the nuclear charge form factor and
$\Gamma_P $ is the two-photon decay width of the pseudoscalar
meson.The differential cross section (\ref{eq:1}) peaks at
$\vec{q}^2 = \Delta^2$, where it rises with the photon's energy
$E_{\gamma}$ as ${d\sigma\over dt} \propto {1/\Delta^2 }
\propto E_{\gamma}^2$, whereas the position of the peak shifts to lower
values of $\vec{q}^2 \propto 1/E_{\gamma}^2$. This property allows
one to isolate the Coulomb contribution unambiguously and measure the
two-photon decay width $ \Gamma_P=\Gamma(P\to 2\gamma)$ of
pseudoscalar mesons. The recent progress of the experimental
technique and high intensity photon beams available at CEBAF
have lead to a new proposal of the measurement of the
neutral pion lifetime at the level of several per mill,
which would allow crucial tests of the chiral anomaly \cite{G}.\\
Subsequently, Pomeranchuk and Shmushkevich \cite{Chuk}
extended the method to the determination of the lifetime of
the $\Sigma^0$-hyperon via Coulomb production of the $\Sigma^0$
in the beam of $\Lambda^0$-hyperons. Subsequently, many radiative
width of many charged meson resonances have been
measured via Coulomb photoproduction of resonances in the
pion and kaon beams (for the review and references see \cite{L}
and the Review of Particle Properties \cite{RPP}).
Because of the C-parity constraints, the Primakoff effect does
not contribute to the photoproduction of the vector mesons, and one
needs at least two-photon exchange. The two-photon exchange contribution
to the photoproduction amplitude will no longer contain the Coulomb pole
inherent to the Primakoff contribution. However, the principal point
that in the Coulomb amplitude the important impact parameters are very large,
\ba
\mid\vec b\mid \sim {1\over \Delta} \label{eq:3}
\ea
remains very much relevant. Furthermore, according to (\ref{eq:2}),
the range of relevant impact parameters rises rapidly with energy.
The Coulomb contribution to the photoproduction amplitude from these
large impact parameters has logarithmic singularity
$\propto \log[1/R_{A}^2(\vec{q}^2+\Delta^2)]$, which generates the
singular dependence of the forward diffraction slope on $\vec{q}^2$ and
energy.Finally, invoking the familiar vector meson dominance, the Coulomb
contribution to the photoproduction amplitude can be related to the
amplitude of elastic scattering of the vector meson in the Coulomb
field. To the two-photon exchange approximation, the latter is
proportional to the radius squared of the vector meson.
Consequently, the experimental isolation of the ultraperipheral
Coulomb contribution to the diffractive vector meson production
would lead to a unique experimental measurement of the size of neutral
vector mesons and must not be overlooked.\\
The estimations of higher order Coulomb corrections have been performed
earlier for photoproduction of pseudoscalar mesons ~\cite{Sch}. There, too,
multiple Coulomb exchanges have the logarithmic singularity, but they are too
weak to produce a numerically substantial correction to the Primakoff
amplitude with the pole singularity.

\section{Color and electric dipole view at the ultraperipheral Coulomb
production and the radius of vector mesons}

Because of the rise of the coherence and formation times, at high
energies the meson photoproduction can be viewed as a three step
process: splitting of the photon into quark-antiquark pair at a
large distance in front of the target, interaction of the
quark-antiquark color and electric dipole with the target, and
projection of the scattered quark-antiquark pair onto the observed
meson. When such a target is a Coulomb field of heavy nuclei, it
is not \'a priori obvious that higher order Coulomb corrections
will be small, because the QED expansion parameter $Z\alpha_{em} \sim 1$.
However, because mesons are electrically neutral and have a
small size $R_M$, the strength of the interaction of small
electric dipole at large impact parameters $\vec{b}$ is suppressed by the
small parameter $R_M^2/\vec{b}^2$, so that the ultraperipheral Coulomb
production can be estimated to the leading two-photon exchange
approximation.\\

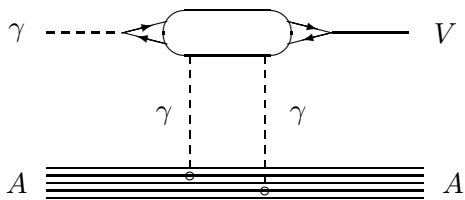
\begin{figure}[!htb]
  \centering
  \setlength{\unitlength}{1cm}
\unitlength=2.0mm \special{em:linewidth 0.8pt}
\linethickness{0.5pt}
\begin{picture}(26.00,15.00)

\put(1.00,1.00){\line(1,0){25.00}}
\put(1.00,1.50){\line(1,0){25.00}}
\put(1.00,2.00){\line(1,0){25.00}}
\put(1.00,2.50){\line(1,0){25.00}}
\put(1.00,3.00){\line(1,0){25.00}}

\put(1.0,12.0){\line(1,0){0.50}} \put(2.0,12.0){\line(1,0){0.50}}
\put(3.0,12.0){\line(1,0){0.50}} \put(4.0,12.0){\line(1,0){0.50}}
\put(5.0,12.0){\line(1,0){0.50}} \put(6.0,12.0){\line(4,1){3.0}}
\put(6.0,12.0){\line(4,-1){3.0}} \put(6.0,12.0){\vector(4,1){2.0}}
\put(9.0,11.25){\vector(-4,1){2.0}}

\put(13.00,12.00){\oval(8.5,3)[1.0]}

\put(20.0,12.0){\line(1,0){5.00}}
\put(20.0,12.0){\line(-4,1){3.0}}
\put(20.0,12.0){\line(-4,-1){3.0}}
\put(20.0,12.0){\vector(-4,-1){2.0}}
\put(17.0,12.75){\vector(4,-1){2.0}}

\put(10.5,10.0){\line(0,1){0.50}} \put(10.5,9.0){\line(0,1){0.50}}
\put(10.5,8.0){\line(0,1){0.50}} \put(10.5,7.0){\line(0,1){0.50}}
\put(10.5,6.0){\line(0,1){0.50}} \put(10.5,5.0){\line(0,1){0.50}}
\put(10.5,4.0){\line(0,1){0.50}} \put(10.5,3.0){\line(0,1){0.50}}
\put(10.5,2.50){\circle{0.60}}

\put(15.5,10.0){\line(0,1){0.50}} \put(15.5,9.0){\line(0,1){0.50}}
\put(15.5,8.0){\line(0,1){0.50}} \put(15.5,7.0){\line(0,1){0.50}}
\put(15.5,6.0){\line(0,1){0.50}} \put(15.5,5.0){\line(0,1){0.50}}
\put(15.5,4.0){\line(0,1){0.50}} \put(15.5,3.0){\line(0,1){0.50}}
\put(15.5,2.0){\line(0,1){0.50}} \put(15.5,1.50){\circle{0.60}}

\put(-1.0,2.0){\makebox(0,0)[cc]{$A$}}
\put(28.0,2.0){\makebox(0,0)[cc]{$A$}}
\put(8.80,6.5){\makebox(0,0)[cc]{$\gamma$}}
\put(17.70,6.5){\makebox(0,0)[cc]{$\gamma$}}
\put(-1.0,12.0){\makebox(0,0)[cc]{$\gamma$}}
\put(27.5,12.0){\makebox(0,0)[cc]{$V$}}
\end{picture}
   \caption{The two-photon exchange contribution to the
photoproduction amplitude}
 \label{fig 1}
  \end{figure}
\begin{figure}[!htb]
  \centering
  \setlength{\unitlength}{1cm}
\unitlength=2.0mm \special{em:linewidth 0.5pt}
\linethickness{0.5pt}
\begin{picture}(80.00,15.00)

\put(1.00,1.00){\line(1,0){25.00}}
\put(1.00,1.50){\line(1,0){25.00}}
\put(1.00,2.00){\line(1,0){25.00}}
\put(1.00,2.50){\line(1,0){25.00}}
\put(1.00,3.00){\line(5,1){12.50}}
\put(26.00,3.00){\line(-5,1){12.50}}

\put(1.0,12.0){\line(1,0){0.50}} \put(2.0,12.0){\line(1,0){0.50}}
\put(3.0,12.0){\line(1,0){0.50}} \put(4.0,12.0){\line(1,0){0.50}}
\put(5.0,12.0){\line(1,0){0.50}} \put(6.0,12.0){\line(4,1){3.0}}
\put(6.0,12.0){\line(4,-1){3.0}} \put(6.0,12.0){\vector(4,1){2.0}}
\put(9.0,11.25){\vector(-4,1){2.0}}

\put(13.00,12.00){\oval(8.5,3)[1.0]}

\put(20.0,12.0){\line(1,0){5.00}}
\put(20.0,12.0){\line(-4,1){3.0}}
\put(20.0,12.0){\line(-4,-1){3.0}}
\put(20.0,12.0){\vector(-4,-1){2.0}}
\put(17.0,12.75){\vector(4,-1){2.0}}

\put(14.0,5.5){\line(-4,1){2.00}} \put(12.0,6.0){\line(4,1){2.00}}
\put(14.0,6.5){\line(-4,1){2.00}} \put(12.0,7.0){\line(4,1){2.00}}
\put(14.0,7.5){\line(-4,1){2.00}} \put(12.0,8.0){\line(4,1){2.00}}
\put(14.0,8.5){\line(-4,1){2.00}} \put(12.0,9.0){\line(4,1){2.00}}
\put(14.0,9.5){\line(-4,1){2.00}}
\put(12.0,10.0){\line(4,1){2.00}}

\put(-1.0,2.0){\makebox(0,0)[cc]{$A$}}
\put(28.0,2.0){\makebox(0,0)[cc]{$A$}}
\put(16.00,8.0){\makebox(0,0)[cc]{${\bf I\!P}$}}
\put(-1.0,12.0){\makebox(0,0)[cc]{$\gamma$}}
\put(27.5,12.0){\makebox(0,0)[cc]{$V$}}

\put(30.5,7.5){\makebox(0,0)[cc]{$+$}}

\put(35.00,1.00){\line(1,0){25.00}}
\put(35.00,1.50){\line(1,0){25.00}}
\put(35.00,2.00){\line(1,0){25.00}}
\put(35.00,2.50){\line(6,1){16.0}}
\put(60.00,3.00){\line(-4,1){9.5}}
\put(35.00,3.00){\line(4,1){10.0}}
\put(60.00,2.50){\line(-6,1){16.0}}

\put(35.0,12.0){\line(1,0){0.50}}
\put(36.0,12.0){\line(1,0){0.50}}
\put(37.0,12.0){\line(1,0){0.50}}
\put(38.0,12.0){\line(1,0){0.50}} \put(38.5,12.0){\line(4,1){3.0}}
\put(38.5,12.0){\line(4,-1){3.0}}
\put(38.5,12.0){\vector(4,1){2.0}}
\put(41.5,11.25){\vector(-4,1){2.0}}

\put(47.00,12.00){\oval(11.5,3)[1.0]}

\put(55.5,12.0){\line(1,0){3.50}}
\put(55.5,12.0){\line(-4,1){3.0}}
\put(55.5,12.0){\line(-4,-1){3.0}}
\put(55.5,12.0){\vector(-4,-1){2.0}}
\put(52.5,12.75){\vector(4,-1){2.0}}
\put(45.0,5.5){\line(-4,1){2.00}} \put(43.0,6.0){\line(4,1){2.00}}
\put(45.0,6.5){\line(-4,1){2.00}} \put(43.0,7.0){\line(4,1){2.00}}
\put(45.0,7.5){\line(-4,1){2.00}} \put(43.0,8.0){\line(4,1){2.00}}
\put(45.0,8.5){\line(-4,1){2.00}} \put(43.0,9.0){\line(4,1){2.00}}
\put(45.0,9.5){\line(-4,1){2.00}}
\put(43.0,10.0){\line(4,1){2.00}}

\put(49.5,5.0){\line(4,1){2.00}} \put(51.5,5.5){\line(-4,1){2.00}}
\put(49.5,6.0){\line(4,1){2.00}} \put(51.5,6.5){\line(-4,1){2.00}}
\put(49.5,7.0){\line(4,1){2.00}} \put(51.5,7.5){\line(-4,1){2.00}}
\put(49.5,8.0){\line(4,1){2.00}} \put(51.5,8.5){\line(-4,1){2.00}}
\put(49.5,9.0){\line(4,1){2.00}} \put(51.5,9.5){\line(-4,1){2.00}}
\put(49.5,10.0){\line(4,1){2.00}}
\put(33.0,2.0){\makebox(0,0)[cc]{$A$}}
\put(62.0,2.0){\makebox(0,0)[cc]{$A$}}
\put(41.50,8.0){\makebox(0,0)[cc]{${\bf I\!P}$}}
\put(53.50,8.0){\makebox(0,0)[cc]{${\bf I\!P}$}}
\put(33.0,12.0){\makebox(0,0)[cc]{$\gamma$}}
\put(61.5,12.0){\makebox(0,0)[cc]{$V$}}
\put(65.5,7.5){\makebox(0,0)[cc]{$+\quad \cdots$}}
\end{picture}
\caption{Left: the single-pomeron exchange, i.e.,
impulse approxi\-ma\-tion, con\-tri\-bu\-tion to the
pho\-topro\-duction off a nucleus,right: the multiple exchange
contribution to the the photoproduction off a nucleus.}
\label{fig 2}
\end{figure}
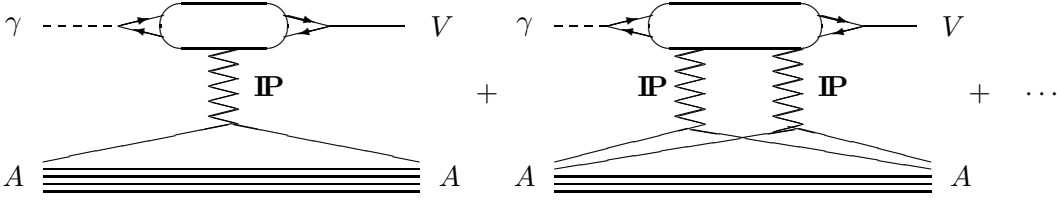

The technical argument why the Coulomb contribution is singular
and measures the size of the vector meson goes as follows.
The vector meson photoproduction amplitude $\gamma +A \to V+A $ is
the sum of Coulomb (Fig.~1), $M_C$, and strong (Fig.~2), $M_s$,
production amplitudes.

As we argued, the Coulomb amplitude can be evaluated to the lowest
order in QED perturbation theory, i.e. the two-photon exchange
approximation. Because the typical impact parameters which contribute to
the Coulomb amplitude are much larger than the size of the vector
meson, $|\vec{b}| \sim 1/\Delta = 2E_{\gamma}/m_V^2 \gg R_{V}$,
the principal quantity is an imaginary part of an amplitude,
$f(\vec{r},\vec{q})$, of a nearly forward scattering of the $q\bar{q}$
electric dipole in the Coulomb field of a nucleus. With some important
modifications for the effect of the finite longitudinal momentum
transfer, which amounts to an effective screening of the Coulomb
potential, the amplitude $f(\vec{r},\vec{q})$ is related to the
electric dipole scattering cross section $\sigma_C(\vec{r})$,
where $\vec{r}$ is the $q\bar{q}$ separation in the
two-dimensional impact parameter space. The latter enters, for instance,
the calculation of a scattering of atoms in a Coulomb field, the case of
muonium (or pionium ~\cite{GTV}) is the most relevant one because the size
of muonium is much smaller than the radius of an atom and nucleus acts
predominantly as a pointlike charge. If we define $f(\vec{r},\vec{q})$ for
the electric dipole made of the particles of charge $\pm 1$, then the
imaginary part of the Coulomb amplitude can be cast in the form
\ba
M_C(\vec{q})= C_C(V)\langle V|f(\vec{r},\vec{q})|\gamma> =
{e \over F_V} \langle V|f(\vec{r},\vec{q})|V\rangle
\label{eq:2.1}
\ea
where the charge-isotopic factors equal
\ba
C_C(\rho) = {1 \over \sqrt{2}}(e_u^3+e_d^3)= {1 \over 3\sqrt{2}}\nonumber\\
C_C(\omega) = {1 \over \sqrt{2}}(e_u^3-e_d^3)= {7 \over 27\sqrt{2}}\nonumber\\
C_C(\phi) = e_s^3 = {1\over 27}
\label{eq:2.1*}
\ea
In (\ref{eq:2.1}) we invoked vector dominance model
(for an extensive review and references see \cite{B}) and
$F_V$ are related to the standard vector dominance constants $f_V$
as
\ba
{1 \over F_{\rho}} = {1\over f_V} (e_u^2- e_u e_d + e_d^2)
= {7 \over 9 f_{\rho}}
\nonumber\\
{1 \over F_{\omega}} = {1\over f_V} (e_u^2+ e_u e_d + e_d^2)
= {1 \over 3 f_{\omega}}
\nonumber\\
{1 \over F_{\phi}} = {1\over f_\phi} e_s^2 = {1\over 9 f_\phi}
\label{eq:2.1**}
\ea
The amplitude (\ref{eq:2.1}) is normalized so that
\ba
\frac{d\sigma}{dq^2}=\frac{{\mid M \mid}^2 }{16\pi}\,.
\label{eq:2.1***}
\ea

In the calculation of the amplitude $f(\vec{r},\vec{q})$ a care must be
taken of the longitudinal momentum transfer. Interaction of the incoming
photon with the first nuclear photon puts the quark-antiquark pair on
mass shell, and the exchanged photon has a longitudinal momentum
\ba
\Delta_1=\frac{M^2}{2E_{\gamma}}\,,
\ea
where $M$ is the invariant mass of the $q\bar{q}$ pair. The second
photon has a longitudinal momentum
\ba
\Delta_2=\frac{M_V^2-M^2}{2E_{\gamma}}\,.
\ea
In the nonrelativistic quark model $M \sim M_V$ and $|\Delta_2| \ll
\Delta_1$. This hierarchy is a good starting approximation
for light mesons as well. In what follows we put
$\Delta_1=\Delta = m_V^2/2E_{\gamma}$ and $\Delta_2 =0$.
For high energy photons the size of the vector meson can
be neglected compared to $1/\Delta$ and
$f(\vec{r},\vec{q})$ can be cast in the form,
\ba
f(\vec{r},\vec{q})=(Z\alpha_{em})^2\vec{r}^2
\int {d^2\vec{k} (\vec{k} - {1\over 2}\vec{q},\vec{k} + {1\over 2}\vec{q})
F_A(\vec{k}-{1\over 2}\vec{q})F_A(\vec{k}+{1\over 2}\vec{q}) \over
[(\vec k-{1\over 2}\vec q)^2+\Delta^2](\vec k+{1\over 2}\vec q)^2} \nonumber\\
=\pi (Z\alpha_{em})^2\vec{r}^2\ln{\frac{1}{(q^2+\Delta^2)\langle R_A^2 \rangle_{ch}}} \,,
\label{eq:8}
\ea
which clearly exhibits the logarithmic singularity of the Coulomb
amplitude as a function of $\vec{q}^2$. Here
$\langle \vec{R}_A^2 \rangle_{ch}$ is the charge radius of the nucleus
squared, and hereafter the capital $\vec{R}$ stands for the 3-dimensional
radius-vector.
The substitution of (\ref{eq:8})
into (\ref{eq:2.1}) gives
\ba
M_C(\vec{q})={\pi e(Z\alpha_{em})^2
\over F_V} \langle V|\vec{r}^2|V\rangle
\ln{\frac{6}{(q^2+\Delta^2)\langle R_A^2 \rangle_{ch}}}\nonumber\\
={8\pi e(Z\alpha_{em})^2
\over 3F_V} \langle \vec{R}_V^2\rangle_{ch}
\ln{\frac{6}{(q^2+\Delta^2)\langle R_A^2 \rangle_{ch}}}
\label{eq:8*}
\ea
where in the case of the $\rho^{\pm}$-meson $\langle \vec{R}_V^2\rangle_{ch}$
is the conventionally defined 3-dimensional
charge radius mean squared.
Consequently, the experimental isolation of the ultraperipheral Coulomb
production amplitude would give a unique possibility to measure the
radius of the vector meson squared.

\section{The amplitude of photoproduction via strong interaction}

Within the QCD color dipole framework, one needs first to
calculate the amplitude of interaction of the $q\bar{q}$ color
dipole with the target, and then use the same formalism as
outlined above for electric dipoles \cite{NNZ}. However, for the
illustration purposes it is more straightforward to invoke the
vector dominance model by which \ba M_s(\vec{q})= {e \over f_V}
M_{VA}(\vec{q})\, . \label{eq:3.1} \ea Anyway, the real
photoproduction of vector mesons is a soft process and
$M_s(\vec{q})$ can not be reliably calculated in the pQCD
framework. The vector meson-nucleus scattering amplitude can be
evaluated in the standard Glauber approximation. For a simple as
yet sufficiently accurate Gaussian parameterization for a nuclear
matter density, \ba n_A(r)=n_0 \exp
\left(-\frac{r^2}{R_A^2}\right) \,, \ea one finds the familiar
expression \ba M_{VA}(\vec{q})&=&A \sigma_{VN}\exp
\left({R_A^2q^2\over{4A}}\right) \sum_{n=1}^A {(A-1)! \over{n\cdot
n!(A-n)!}}\nn &&\times
\left[-\frac{\sigma(VN)}{2\pi(R_A^2+2B_N)}\right]^{n-1} \exp
\left(-\frac{(R_A^2+2B_N)q^2}{4n} \right)\, , \label{eq:3.3} \ea
where $B_N$ is the diffraction slope of the $VN$ elastic
scattering. In (\ref{eq:3.3}) we neglected the small real part of
the VN scattering amplitude. The total cross section of the vector
meson interaction with nucleon $\sigma(VN)$ can be obtained from
the relation
$$\sigma(VN)=\frac{\sigma(\pi^+N)+\sigma(\pi^-N)}{2}\,.$$
For the estimation purposes, one can neglect $B_N$ compared to
$R_A^2$ and use the simple parameterization
\ba
M_{VA}(\vec{q}) \approx \sigma(VN)A^{\alpha}\exp\left[-{1\over 2}
B_A\vec{q}^2\right]
\label{eq:3.4}
\ea
where for the $\rho$ and $\omega$ the exponent $\alpha \approx 0.8$
and to the crude approximation $B_A \approx {1\over 3} \langle
R_A^2\rangle_{ch}$.

\section{Is the Coulomb correction observable?}

The Coulomb correction to the differential cross section is small,
\ba
\frac{2M_C}{M_s}= \eta_C
\ln{\frac{6}{(\vec{q}^2+\Delta^2)\langle \vec{R}_A^2 \rangle_{ch}}}
\nonumber \\
={16\pi (Z\alpha_{em})^2 f_V
\over 3F_V A^{\alpha}}\cdot {\langle \vec{R}_V^2\rangle_{ch} \over
\sigma(VM)}\cdot
\ln{\frac{6}{(\vec{q}^2+\Delta^2)\langle \vec{R}_A^2 \rangle_{ch}}}
\,.
\label{eq:ratio}
\ea
Take the photoproduction of the $\rho$ meson off the lead target, for
which $\langle \vec{R}_A^2 \rangle_{ch} \approx 25$ f$^2$.
The nonrelativistic quark model suggests
$\langle \vec{R}_V^2\rangle_{ch} \approx \langle \vec{R}_{\pi}^2\rangle_{ch}
\approx 0.4$ f$^2$ and $\sigma(\rho N)\sim 30 mb$. For $E_{\gamma}=
10 GeV$ the logarithmic factor is of the order of 2, so that
the Coulomb correction is negligible small, $\sim 1.5 \%$.

The logarithmic singularity of $2M_C(\vec{q})$ makes the Coulomb
correction much more noticeable in the diffraction slope
\ba
B=2\frac{d\ln M}{d\vec{q}^2}
\ea
Using the results (15), (17), (18), (22) one obtains for the
slope of the
full amplitude
\ba
B=B_A+\Delta B_C=B_A + \eta_C {1\over \vec{q}^2+\Delta^2} =
B_A + {4\eta_C E_{\gamma}^2 \over m_{\rho}^4}\cdot{\Delta^2
\over \vec{q}^2+\Delta^2}
\,
\label{eq:slope}.
\ea

In order to emphasize the importance of the  Coulomb correction
to the diffraction slope, $\Delta B_C$, start with the
exactly forward, $\vec{q}^2=0$, photoproduction off the lead target.
Here teh Coulomb correction rises from $\Delta B_{C} \sim 20$ GeV$^{-2}$
at  $E_{\gamma}=10$ GeV to a dramatically large
$\Delta B_{C} \sim 2000$ GeV$^{-2}$ to be compared to $B_A \sim 200$
GeV$^-2$ at $E_{\gamma}=100$ GeV. Of course, this dramatic rise is
confined to a region of
ultrasmall momentum transfers, from $|\vec{q}| \sim \Delta \approx 2.5
{\rm MeV/c}$ at $E_{\gamma}=100$ GeV, which are perhaps beyond the
experimental reach, to a more realistic
$|\vec{q}| \sim \Delta \approx 25 {\rm MeV/c}$ at $E_{\gamma}=10$ GeV.
Now notice, that while the prefactor $\eta_C$ does not depend on
our approximations for $\Delta_{1,2}$, the exact $\vec{q}^2$ dependence,
and the magnitude at $\vec{q}^2=0$, of the $\Delta B_C$, do explicitly
depend on $\Delta_{1,2}$. Consequntly, for the accurate determination
of the size of the $\rho$-meson one must concentrate on the
Coulomb correction $\Delta B_C$ at  $\vec{q}^2 >\Delta^2$, where the
$\vec{q}^2$ dependence is insensitive to $\Delta_{1,2}$.

\section{Summary and conclusions}

We pointed out a new, ultraperipheral Coulomb, mechanims for the
photoproduction of vector mesons off nuclei at ultrasmall momentum
transfer. The Coulomb contribution to the photoproduction
amplitude has a singular, logarithmic, dependence on the
momentum transfer squared, and is proportional to
the mean radius of the vector veson squared. This strongly energy dependent
singular contribution entails a steep dependence of the diffraction
slope on $\vec{q}^2$.
The very specific dependence of the Coulomb correction $\Delta B_C$
on the photon energy, the momentum transfer squared $\vec{q}^2$, the
target charge and mass number makes the experimental evaluation of
the size of the $\rho$ meson quite plausible.
The work of I.I. and S.G. has been partly supported by the
grants INTAS 00-0036 and INTAS 97-30494.

\end{document}